# Optical and Microcavity Modes Entanglement by means of Opto-Electronics System

Ahmad Salmanogli[1*], Dincer Gokçen[2], H. Selcuk Geçim[1]

[1]Faculty of Engineering, Electrical and Electronics Engineering Department, Çankaya University, Ankara, Turkey
[2] Electrical and Electronics Engineering Department, Hacettepe University, 06800, Ankara, Turkey.

**Abstract:** Entanglement between optical mode and microwave mode is a critical issue in illumination systems. Traditionally, optomechanical systems are applied to couple the optical mode to microcavity modes. However, due to some restrictions of this system such as sensitivity to the thermal noise at room temperature, a novel optoelectronic system is designed in this study to fix the recent system problems. Unlike recent optomechanical systems, the optical modes are directly coupled to the microwave cavity through the optoelectronic elements with no need mechanical system. The main objective of this work is to generate the entangled modes at room temperature. For this purpose, we theoretically analyze the dynamics of motion of the optoelectronic system with the Heisenberg-Langevin equations, from which one can calculate the coupling between optical and microcavity modes. The most important feature of this system is the coupling between optical mode and microwave cavity mode, which is done using a photodetector and a Varactor diode. Hence, by controlling the photodetector current (photocurrent), which is dependent on the optical cavity incidence wave and the Varactor diode biased, which is the function of the photocurrent, the coupling between the optical and microcavity mode is fully done. The modeling results show that the coupled modes are entangled at the room temperature with no need to any mechanical parts.

## I. INTRODUCTION

In the recent years, fabrication of the sensitive sensors has been a subject of intense research. Undoubtedly, the technological trend would replace the classical sensors by the quantum ones because the quantum-based sensors are able to induce some important advantages to improve the sensory performance such as resolution enhancement via quantum imaging [1, 2], quantum performance effect on sensitivity improving in quantum radar [3, 4], and so on. Actually, quantum-based sensors produce nonclassicality and entanglement in various approaches [5-7] and utilized these novel phenomena to improve performance in different applications such as plasmonic [8, 9], Raman spectroscopy [10], and other cases. Generally speaking, entanglement occurs when two photons are generated to interact in such a way that their properties are linked together [11, 12]. It has to be noted that the link between the related photons or states is nonclassical and does not obey the classical laws. Accordingly, determining the position, momentum, spin, or polarization of one photon or state at a place, not dependent on the inter-distance between the photons, allows examining the complementary position, momentum, spin, or polarization of their partner at another place [3]. To date, this unique quantum property has been widely used in different applications to improve the system specifications, which are not accessible in the case of the classical system. This system that utilizes the entanglement to enhance the system features is indeed a quantum illumination system [3], [4]. Also, for generating an entangled state, which is the base of this type of studies, mainly the optomechanical system is used. This system generates the entanglement between optical cavity (OC) and microwave (MW) cavity [13-17]. In this system, which uses the mechanical part, some photos generated due to the thermal effect in high temperature restrict the system's operation at high temperature. This problem is induced because the mechanical part operation frequency is very low compared to MW cavity and OC. Indeed, the thermal photon numbers generated in the mechanical sub-system limits the system. Some engineering works have been conducted on the mechanic bandwidth engineering to fix this problem [14-16]. In fact, without the mechanical's bandwidth or sideband engineering, the thermal photons strongly restrict the entanglement between cavity modes. Furthermore, in our latest work [17], we defined a new system that operates based on the plasmonic effect; with employing this effect, which is transferred via the capacitor embedded into the optical cavity, a high operational temperature (100 ~ 150 K) was accessible. Both bandwidth engineering and manipulating optical cavity modes may solve slightly the problem; however, the number of the thermal photon generated at room temperature is so high that restricts the cavity modes entangling. In this paper, we will completely fix this problem by canceling out the mechanical part. Indeed, we add an optoelectronic subsystem by which the OC modes are easily coupled to the MW cavity modes with no need to any mechanical elements. In the present system, the OC modes are initially coupled to a photodetector, which leads to generating a contributed photocurrent in the near-infrared region (NIR) [18-20]; this photocurrent biases the Varactor diode [21], suggesting that this diode bias voltage is a function of the incoming light from the OC. Moreover, it is obvious that the Varactor diode capacitance varies with its bias voltage alteration. In other words, the Varactor diode capacitance alters with OC coupling mode alteration; hence, the LC circuit frequency is a function of the Varactor capacitance changing, which is modified due to the OC coupling to the photodiode. Actually, we prove and illustrate that using optoelectronic system enables cavity modes to remain entangled even at room temperature with a slight influence from the thermal photon generating in the system. Using this system, the induced complexity of the optomechanical system is dramatically reduced and the degree of freedom for engineering the cavity modes entangling is increased. The study's theory and system description are completely discussed in what follows.

## II. THEORY AND BACKGROUNDS
### A. System description:

The proposed optoelectronic system is illustrated with a sufficient detail in Fig. 1. Fig. 1a shows the system's schematic, which consists of an optical cavity, a photodetector, Varactor diode, and inductor. In this system, it is schematically observed that the optical cavity mode is initially coupled to the photodetector, indicating that the photocurrent is a function of the coupling light from the optical cavity, $i=f_1(\hbar\omega)$ where $f_1$ is a coupling factor function. This current flows through the Varactor diode and so the Varactor diode bias voltage depends on the photocurrent. Thus, Varactor diode voltage, which handles the diode-capacitor, is strongly dependent on the optical cavity incident light, $C=f_2(\hbar\omega)$, where $f_2$ is a coupling factor function. Hence, the microwave cavity output frequency is highly attributed to the Varactor diode capacitor, which is a function of the photocurrent. Such optoelectronic design satisfies a direct coupling between the optical cavity and the microwave cavity fields.

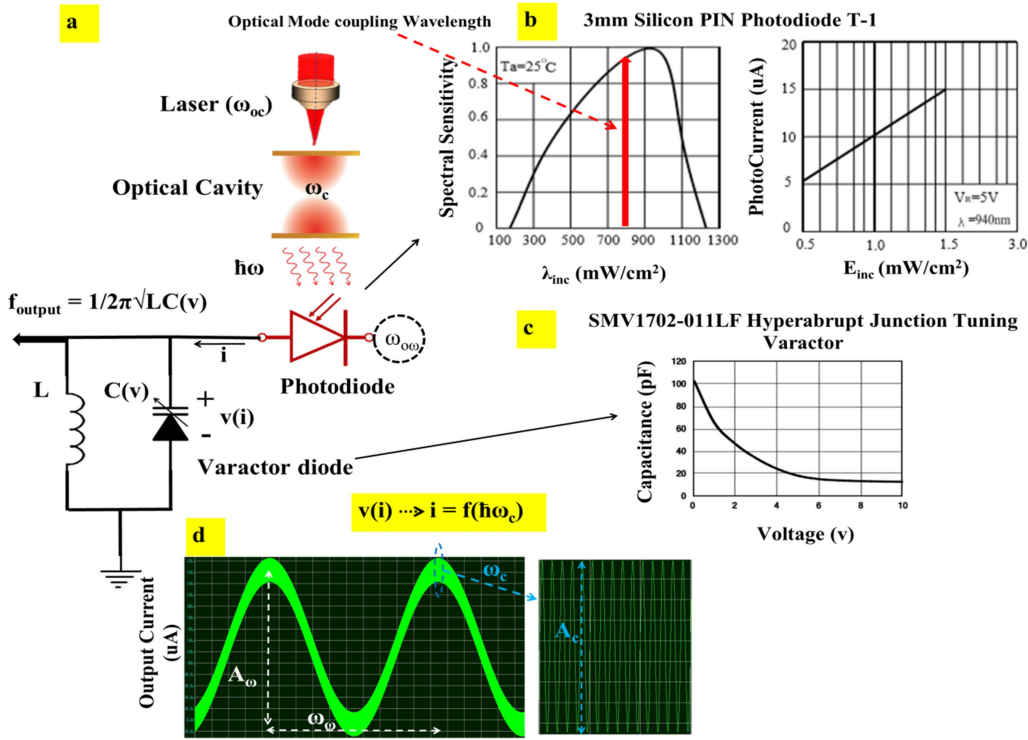

Fig. 1 Optoelectronic system schematic, a) OC mode coupling to MW cavity mode through the photodetector and a Varactor diode, b) a typical photodiode spectral sensitivity and photocurrent graphs [28], c) Varactor diode capacitance variation vs biased voltage [29], d) simulated photocurrent as a function of optical cavity mode incidence wave frequency and amplitude.

Moreover, in this figure, we show the typical model for the photodetector illustrated in Fig. 1b. This figure shows the photodetector spectral sensitivity and photocurrent. In Fig. 1c, the Varactor diode capacitance changing is shown versus its biased voltage; in the same way, a typical model was selected for this diode. The illustrated models are just to give a sense to anyone in the case of photodetector current, sensitivity, and Varactor diode operation; and actually they are used only for comparison with the present study's modeling results. However, a more important case, which was simulated in PROTEUS software, is illustrated in Fig. 1d. This figure shows the coupling effect of the optical cavity and, also, how it is coupled to the MW cavity modes, where $A_\omega$ and $\omega_\omega$ are the MW cavity amplitude and frequency, respectively. Furthermore, it shows the OC mode frequency ($\omega_c$) and amplitude ($A_c$), and one can surely find that $\omega_c \gg \omega_\omega$. It is important to remark that by engineering the $f_1$ and $f_2$, the coupling between two cavity modes will be handled. It means that the ratio of the $f_1/f_2$ will be translated to the ratio of the $A_c/A_\omega$, which defines the effect of the optical cavity mode on the microcavity mode. Indeed, we want to prove that by ratio engineering, the cavity modes entangling is accessible. In the following, the theory and background section of this work are presented.

*B. Optoelectronic system dynamics of motions*

The total Hamiltonian of the illustrated optoelectronic system in Fig. 1a is defined as [13, 17]:

$$H_{total} = \frac{\varphi^2}{2L} + \frac{Q^2}{2C} - e(t)Q + \hbar\omega_c a^+ a - \hbar G_{oc} a^+ a \\ + i\hbar E_c(a^+ e^{(-j\omega_{oc}t)} - a e^{(j\omega_{oc}t)}) \quad (1)$$

where ($\varphi$, Q) are the canonical coordinates for MW cavity, $\varphi$, L, and Q are the inductor flux, inductance in Henrys, and charge on the capacitor C, respectively. Moreover, a and $a^+$ are the annihilation and creation operator for OC, which is pumped with frequency $\omega_{oc}$. Finally, $E_c$, e(t), and $G_{oc}$ are the input driving laser, MW cavity driving, and coupling factor between OC and photodiode which involves the MC modes too, respectively. Notably, $G_{oc}$ manages the ratio between $f_1$ and $f_2$. By defining the lowering and raising operator for MW cavity and by surpassing the fast oscillating parts, the system's total Hamiltonian is introduced as:

$$H_{total} = \hbar\Delta_\omega b^+ b + \hbar\Delta_c a^+ a - \hbar q_{oc} a^+ a b^+ b \\ + i\hbar E_c(a^+ - a) + i\hbar E_\omega(b^+ - b) \quad (2)$$

where $\Delta_\omega = \omega_\omega - \omega_{0\omega}$ and $\Delta_c = \omega_c - \omega_{0c}$, and $q_{oc}$ is proportional to $G_{oc}$. Notably, the third term in total Hamiltonian in Eq. 2 defines the interaction Hamiltonian, and more specifically, introduces the interaction of the OC mode intensity with the photodetector and then applying a nonlinear process managed by $q_{oc}$ leads to converting the contributed OC modes to MW cavity photon number. However, the dynamics of the contributed modes in this system are also affected by the damping and noise operator since each mode interacts with its own environment. Hence, the equations of motions for the presented system using Heisenberg-Langevin equation are introduced as:

$$\dot{a} = -(i\Delta_c + \kappa_c)a + iq_{oc}ab^+b + E_c + \sqrt{2\kappa_c}a_{in} \quad (3)$$
$$\dot{b} = -(i\Delta_\omega + \kappa_\omega)b + iq_{oc}ba^+a + E_\omega + \sqrt{2\kappa_\omega}b_{in}$$

where $a_{in}$ and $b_{in}$ are the OC cavity input noise, and MW cavity input noise, respectively. Also, $\kappa_c$ and $\kappa_w$ indicate the OC and MW cavity decay rate, respectively. It is clear that Eq. 3 is a nonlinear equation. To linearize this equation, one can use the fluctuations associated with the field modes as $q = \langle q \rangle + \delta q$, where $\langle q \rangle$ stands for field average in the steady-state condition and $\delta q$ presents the fluctuation of the considered mode [14, 17]. We can certify the linearized approach because of the following reasons. (i) We do not linearize the field itself while we put a $\langle q \rangle$ and solve the system exactly and numerically and (ii) Afterward, we linearize the fluctuations that are noise, not the system itself, around the $\langle q \rangle$. Linearization of the noise in the system can be performed whenever the nonlinearities altering the noise are small. So, it is possible to obtain the equations of motion based on the fluctuation in the form of linearization approximation as:

$$\dot{\delta a} = -(i\Delta_c + \kappa_c)\delta a + iq_{oc}(|\beta|^2 \delta a + \beta\alpha\delta b^+ + \beta^*\alpha\delta b) + \sqrt{2\kappa_c}\delta a_{in}$$
$$\dot{\delta b} = -(i\Delta_\omega + \kappa_\omega)\delta b + iq_{oc}(|\alpha|^2 \delta b + \beta\alpha\delta a^+ + \beta\alpha^*\delta a) + \sqrt{2\kappa_\omega}\delta b_{in}$$
(4)

where $\alpha = \langle a \rangle$ and $\beta = \langle b \rangle$ are the OC and MW cavity average modes, respectively. These parameters are estimated through the steady state equations introduced as:

$$-(i\Delta_c + \kappa_c)\alpha + iq_{oc}\alpha|\beta|^2 + E_c = 0 \quad (5)$$
$$-(i\Delta_\omega + \kappa_\omega)\beta + iq_{oc}\beta|\alpha|^2 + E_\omega = 0$$

we can assume that $|\alpha|$ and $|\beta| \gg 1$ [13]; therefore, $\alpha$ and $\beta$ are approximated as:

$$\alpha = \frac{E_c}{\kappa_c + i(\Delta_c - q_{oc}|\beta|^2)}, \quad \beta = \frac{E_\omega}{\kappa_\omega + i(\Delta_\omega - q_{oc}|\alpha|^2)} \quad (6)$$

By solving $\alpha$ and $\beta$, which are the expectation value of the OC and MW cavity modes, respectively, Eq. 4 is obtained. To study the entanglement between two cavities mode, one need to quadrature fluctuation, so here beside of $\delta a$ and $\delta b$ in Eq. 4, we need to calculate their complex conjugate as $\delta a^+$ and $\delta b^+$. Finally, the general form of Eq. 4 with considering the operator's conjugates can be presented as:

$$\begin{bmatrix} \dot{\delta a} \\ \dot{\delta a}^+ \\ \dot{\delta b} \\ \dot{\delta b}^+ \end{bmatrix} = \underbrace{\begin{bmatrix} -\gamma_c + iq_{oc}|\beta|^2 & 0 & i\alpha\beta q_{oc} & i\alpha\beta^* q_{oc} \\ 0 & -\gamma_c^* - iq_{oc}^*|\beta|^2 & -i\alpha^*\beta^* q_{oc}^* & -i\alpha^*\beta q_{oc}^* \\ i\alpha^*\beta q_{oc} & i\alpha\beta q_{oc} & -\gamma_\omega + iq_{oc}|\alpha|^2 & 0 \\ -i\alpha^*\beta^* q_{oc}^* & -i\alpha\beta^* q_{oc}^* & 0 & -\gamma_\omega^* - iq_{oc}^*|\alpha|^2 \end{bmatrix}}_{A_{i,j}} \times \underbrace{\begin{bmatrix} \delta a \\ \delta a^+ \\ \delta b \\ \delta b^+ \end{bmatrix}}_{u(t)} + \underbrace{\begin{bmatrix} \sqrt{2\kappa_c}\delta a_{in} \\ \sqrt{2\kappa_c}\delta a_{in}^+ \\ \sqrt{2\kappa_\omega}\delta b_{in} \\ \sqrt{2\kappa_\omega}\delta b_{in}^+ \end{bmatrix}}_{n(t)} \quad (7)$$

where $\gamma_c = \kappa_c + i\Delta_c$ and $\gamma_\omega = \kappa_\omega + i\Delta_\omega$. Eq. 7 is solved as $u(t) = \exp(A_{i,j}t)u(0) + \int(\exp(A_{i,j}s).n(t-s))ds$, where $n(s)$ is the noise column matrix. The contributed input noises obey the following correlation function [9-11].

$$\begin{aligned} &<a_{in}(s)a_{in}^*(s')> = [N(\omega_c)+1]\delta(s-s') \\ &<a_{in}^*(s)a_{in}(s')> = [N(\omega_c)]\delta(s-s') \\ &<b_{in}(s)b_{in}^*(s')> = [N(\omega_\omega)+1]\delta(s-s') \\ &<b_{in}^*(s)b_{in}(s')> = [N(\omega_\omega)]\delta(s-s') \end{aligned} \quad (8)$$

where $N(\omega) = [\exp(\hbar\omega/k_B T)-1]^{-1}$ and $k_B$ and $T$ stand for Boltzmann's constant and operation temperature, respectively [9, 11]. Indeed, $N(\omega)$ is the equilibrium mean thermal photon numbers of the different modes. Another interesting point is the optoelectronic system stability. In this regard, one can examine the $A_{ij}$ Eigenvalues to identify the system stability, having the knowledge that for a stable system all of the real parts of the contributed Eigenvalue should be negative [13, 14]. Hence, we are ready to analyze the quantum features of the presented system. Since we tend to study the entanglement between modes in the developed system, we shall focus on the OC-MW modes. With knowledge of the various literature, such bipartite entanglement is quantified through the Symplectic eigenvalue introduced as [11- 12]:

$$\eta = \frac{1}{\sqrt{2}}\sqrt{\sigma \pm \sqrt{\sigma^2 - 2\det(\sigma)}}, \sigma = \det(A) + \det(B) - 2\det(C) \quad (9)$$

where A, B, and C are 2×2 correlation matrix elements [A, C; $C^T$, D]. It should be noted that Eq. 9 is an important criterion for identification of the entanglement between two modes. Based on this equation, it was found that if $2\eta > 1$, the considered modes are purely separable; otherwise, for $2\eta < 1$ two modes are entangled [12]. In the present study, we used this criterion to determine whether the considered modes are entangled. In the next section, the key parameter used in the above equations, which are the OC and MW cavity coupling factor ($q_{oc}$), are analyzed. In fact, it is an important factor by which one can control the coupling between two cavities and, eventually, examine the entanglement between modes. This factor in the present study leads to generating photocurrent through the coupling of the OC to the photodiode ($f_1$), and then the flowing current strongly influences the Varactor biased voltage ($f_2$). Moreover, it has to be noted that the generated photocurrent dramatically depends on the environment temperature, which is studied in the following.

### C. Photocurrent calculation using the time-dependent perturbation theory

To calculate the photodetector's flowing current due to the incidence wave, we applied the time-dependent perturbation theory because the light field shone on materials is relatively weak compared to the fields within the materials. By this theory, we calculate the wave function at some later time; accordingly, we need specifically to know about the system Hamiltonian as $H = H_0 + H_{int}(t)$; to deal with such a situation, we use the time-dependent Schrödinger equation $i\hbar d|\Psi>/dt = H|\Psi>$, where $|\Psi>$ is the time-varying wave vector. With supposing $|\Psi_n>$ and $E_n$ respectively as the Eigenfunction and Eigenvalue of the time-independent equation, it is known that $H_0|\Psi_n> = E_n|\Psi_n>$; therefore, one can expand the solution of the time-dependent equation as $|\Psi> = \{\sum_n a(t)\exp[-iE_n t/\hbar]\}|\Psi_n>$. In the following, by substituting $|\Psi>$ in the time-dependent Schrödinger equation and by pre-multiplying in $<\Psi_q|$, and also by considering the perturbation series, the first-order Eigen function [22- 23] is introduced as:

$$\dot{a}_q^{(1)}(t) = \frac{1}{i\hbar}\cdot\sum_n a_n^{(0)}(t).e^{(i\omega_{qn}t)} <\Psi_q|H_{int}(t)|\Psi_n>, \quad (10)$$
$$\omega_{qn} = \frac{E_q - E_n}{\hbar}$$

where $|\Psi_n>$ is the initial state and $|\Psi_q>$ is the state which we are interested in (a favorite state). To solve Eq. 10, one has to know about the perturbing Hamiltonian for the presented system. For the sake of simplicity, we define the perturbing Hamiltonian semiclassically as $H_{int}(t) = -q.E(t)$; hence, by substituting interaction Hamiltonian in Eq. 10 and with extra assumption that the perturbation starts at t = 0 and ends at t = $t_0$, $a_q^{(1)}(t)$ becomes as:

$$a_q^{(1)}(t) = \frac{1}{i\hbar}\cdot\int_0^{t_0} e^{(i\omega_{qn}t)} <\Psi_q|H_{int}(t)|\Psi_g>dt$$
$$= \frac{1}{i\hbar}\cdot<\Psi_q|H_{int0}|\Psi_g>\int_0^{t_0} e^{(i[\omega_{qn}-\omega]t)} + e^{(i[\omega_{qn}+\omega]t)}dt \quad (11)$$

In this system, we just look at the OC modes coupling to a photodiode, which is the photodiode absorption probability; so, from Eq. 11, the total rate of the transition in the presence of the OC electrical field is introduced as:

$$q_{oc} = \frac{2\pi}{\hbar}\cdot|<\Psi_q|H_{int0}|\Psi_g>|^2 L(\omega_{qg}).g_J(\hbar\omega_{qg}) \quad (12)$$

where $L(\omega_{qg})$ is the Lorentzian function. Actually, we use this function because no line becomes arbitrarily sharp, suggesting that its energy is well defined. Also, we have to introduce a very dense transition and $g_J(\omega_{qg})$ is considered for this aim. Finally, by considering the interaction Hamiltonian and using the Fermi's Golden rule [24, 25], the coupling factor, which is also the total transition rate, is given by:

$$q_{oc} = \frac{2\pi}{\hbar} \cdot \frac{e^2 A_0^2}{4 m_0^2} \cdot |P_{cv}|^2 \cdot \frac{1}{2\pi^2} \cdot \left(\frac{2\mu_{eff}}{\hbar^2}\right)^{1.5} \\ \cdot \sqrt{\hbar\omega_c - E_{gap}} L(\omega_{eg}) \cdot f(T) \quad (13)$$

where e, $m_0$, $A_0^2$, $P_{cv}$, $\mu_{eff}$, and $E_{gap}$ are, respectively, the electron charge, free electron mass, optical intensity, and matrix element that equals <c|**P**|v>, where c and v denote, respectively, the conduction and valence band, electron-hole effective mass, photodetector energy bandgap. More importantly, f(T) in this formula is the temperature factor that depends on the thermal distribution of the electron in excited and ground states, and also relaxation time from the photoexcited state to other states. Generally, this function is the main reason for the strong dependence of the photocurrent on the operational temperature, which means that at the high temperature the photocurrent is dramatically decreased [26, 27]. In the following, we calculate the photocurrent from the total transition rate and study the effect of the environment temperature on this parameter. Eq. 13 introduces the total transition rate, which is the actual coupling parameter between OC and MW cavity modes. From Fig. 1d, we defined the $A_c$ and $\omega_c$ as the photocurrent oscillation amplitude and frequency, which contribute to the OC modes coupling to the photodiode. Indeed, by engineering the $q_{oc}$, one can manipulate the $A_c$ and $\omega_c$; so, $q_{oc}$ is a sole parameter that can be used to control the OC and MW cavity modes. It is noteworthy that $q_{oc}$ is dramatically affected by the OC modes frequency, and also by the environment temperature.

### III. RESULTS AND DISCUSSION

Due to the role of the transition rate factor in the case of the coupling between two modes and its effect on the quantum features of the attributed modes, we studied the parameters that influence the photocurrent; i.e., environment temperature and OC modes frequency. It was supposed that this photodetector is a Si photodetector with energy bandgap around 1.1 eV. Hence, by sweeping the OC frequencies, the amount of the generated photocurrent is modeled and recorded. It should be noted that by increasing the operational temperature, the photocurrent amplitude is decreased because of the thermal noise effect on the photodetector operation. Moreover, the maximum amplitude for each graph occurs where the optical cavity mode incidence frequency is close to the Si bandgap energy. Furthermore, the dark current effect, for simplicity, is neglected.

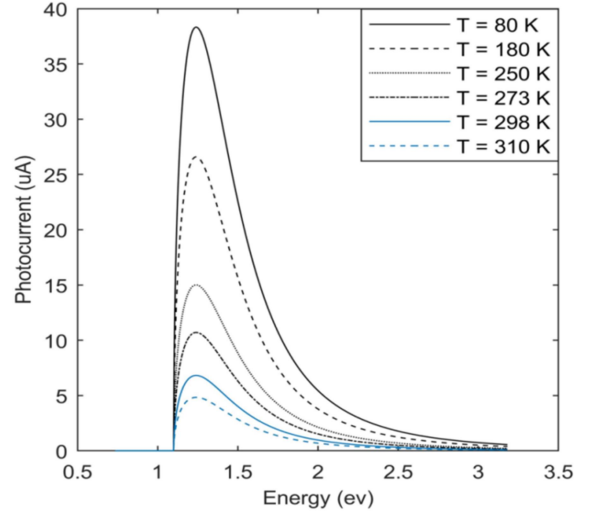

Fig. 2 Photocurrent (μA) vs optical cavity wave energy (ev) at different temperatures, $P_c = P_w = 10$ mW.

The results of the optoelectronic circuit parameters influence on the entanglement between OC and MW cavity modes are illustrated in Fig. 3. The influence of the temperature is considered in the same way of calculating the photocurrent in Fig. 2. Generally, we know that by increasing the operating temperature the entanglement between two modes should be decreased. Thus, one can compare the amplitude of the sub-figures in Fig. 3. It is shown that by increasing the temperature, 2η amplitude is generally increased, meaning that the separability between modes is increased. However, it should be noted that this is correct for every MW cavity detuning frequency $\Delta_\omega/\omega$ (shown with a red-dashed circle) except at $\Delta_\omega/\omega = 0$. For better understanding, we can compare the red-dashed circle in sub-figures; for example, the comparison between Fig. 3a with Fig. 3e reveals that at T = 80 for many MW cavity detuning frequencies, two cavity modes remained entangled whereas at T= 298 this case is dramatically reduced, which is strongly contributed to the thermal photon generated in the system. Moreover, the divergence slopes at each profile are excessed when the temperature is increased. Actually, an interesting event occurs where the $\Delta_\omega/\omega$ equals zero and was indicated with the blue-dashed circle on a few figures. In this regard, it is shown that by increasing the temperature the two considered modes remain entangled. This result is the most important achievement that we got from this study, while in our later work [17] we theoretically proved that by a simple optomechanical system without mechanical bandwidth engineering, the room temperature entanglement is not accessible.

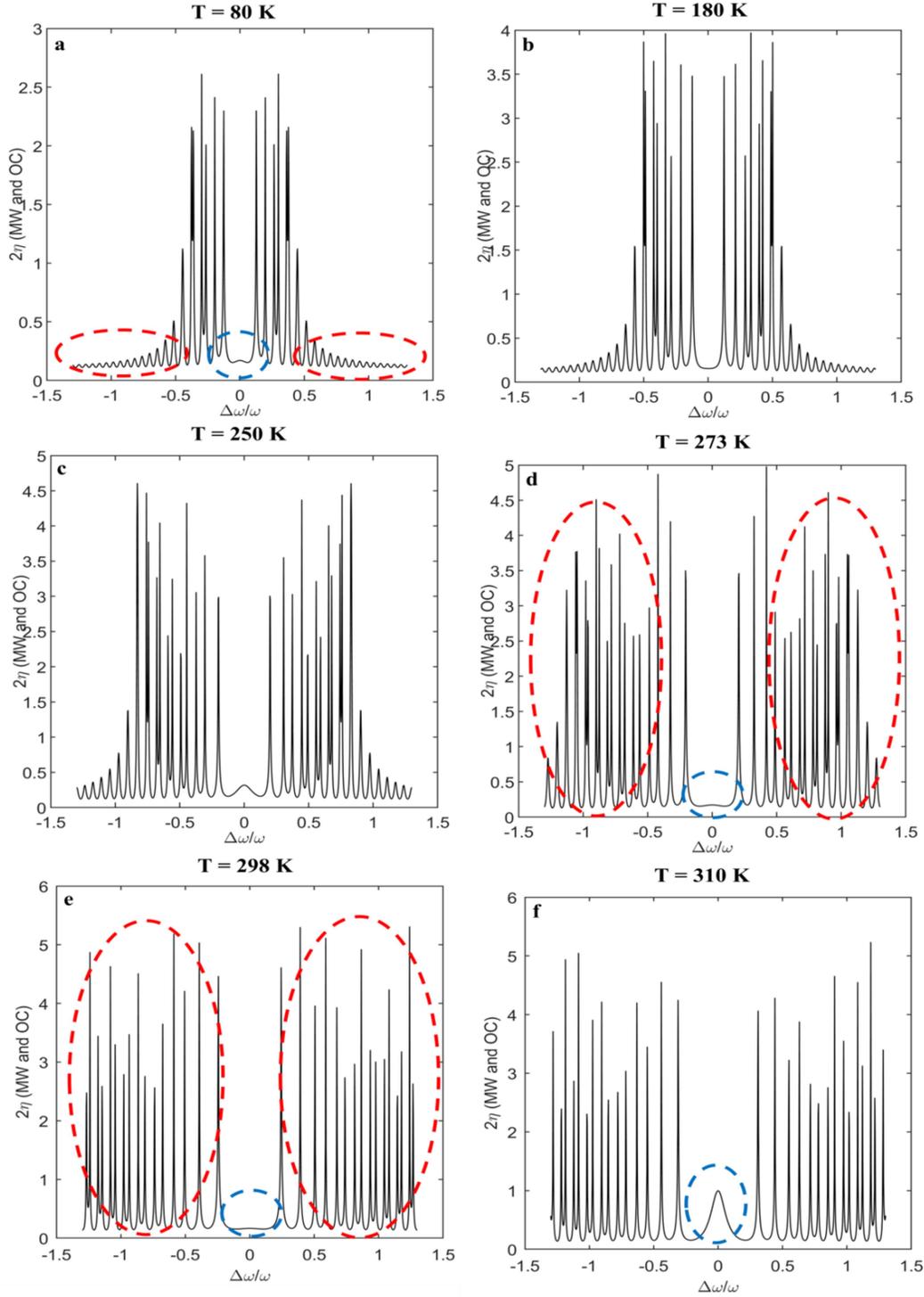

Fig. 3 Entanglement analyzing for two (MW and OC) cavity modes vs $\Delta\omega/\omega$ at $\Delta c = 0$ at different temperatures, a) 80 K, b) 180 K, c) 250 K, d) 273 K, e) 298 K, f) 310 K; , $P_c = P_w = 10$ mW.

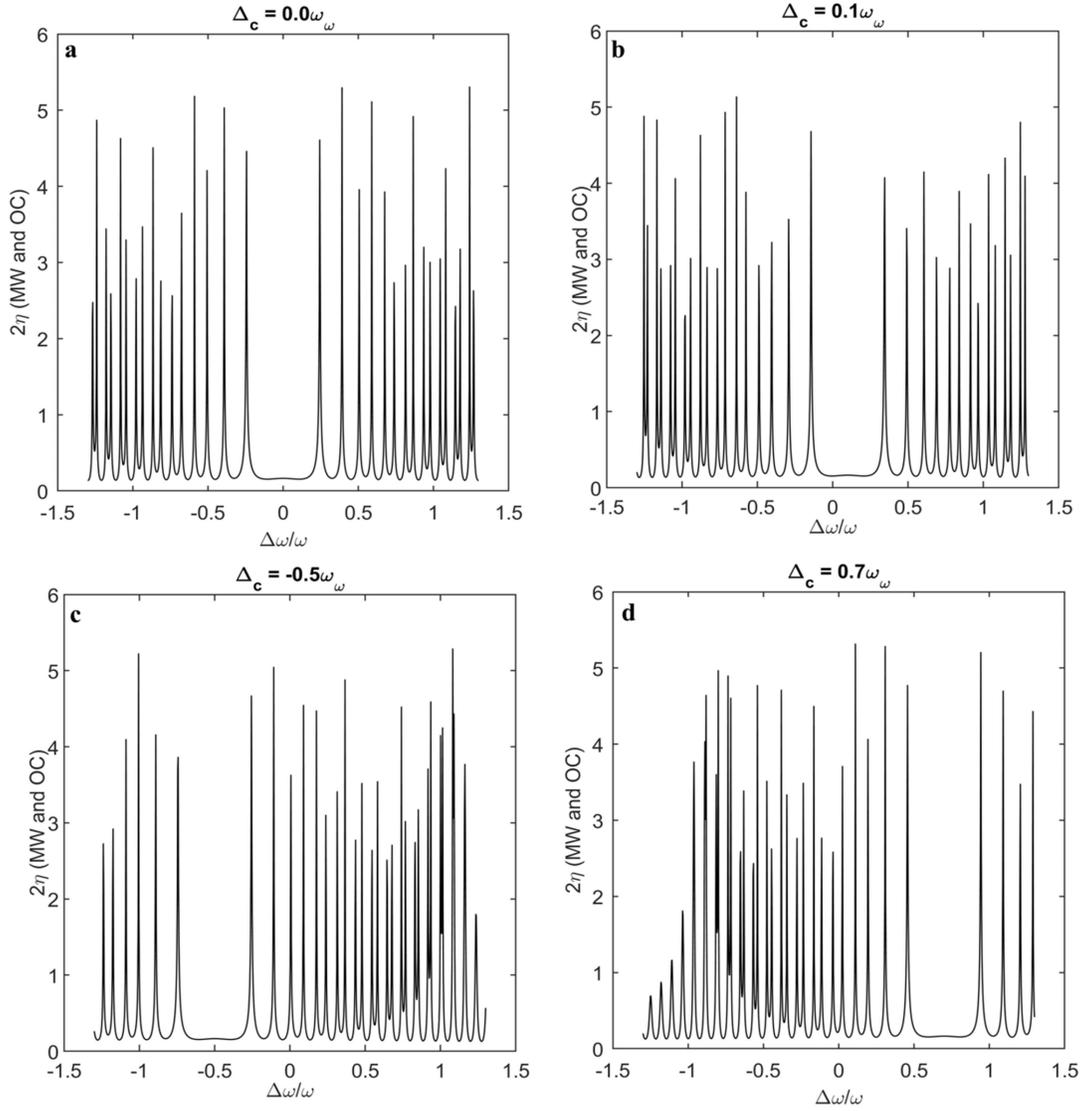

Fig. 4 Entanglement analyzing for two (MW and OC) cavity modes vs $\Delta\omega/\omega$ at different $\Delta_c$, a) $\Delta_c= 0\omega_\omega$, b) $\Delta_c= 0.1\omega_\omega$, c) $\Delta_c= -0.5\omega_\omega$, d) $\Delta_c= 0.7\omega_\omega$; T = 298 K, $P_c = P_w$ = 10 mW.

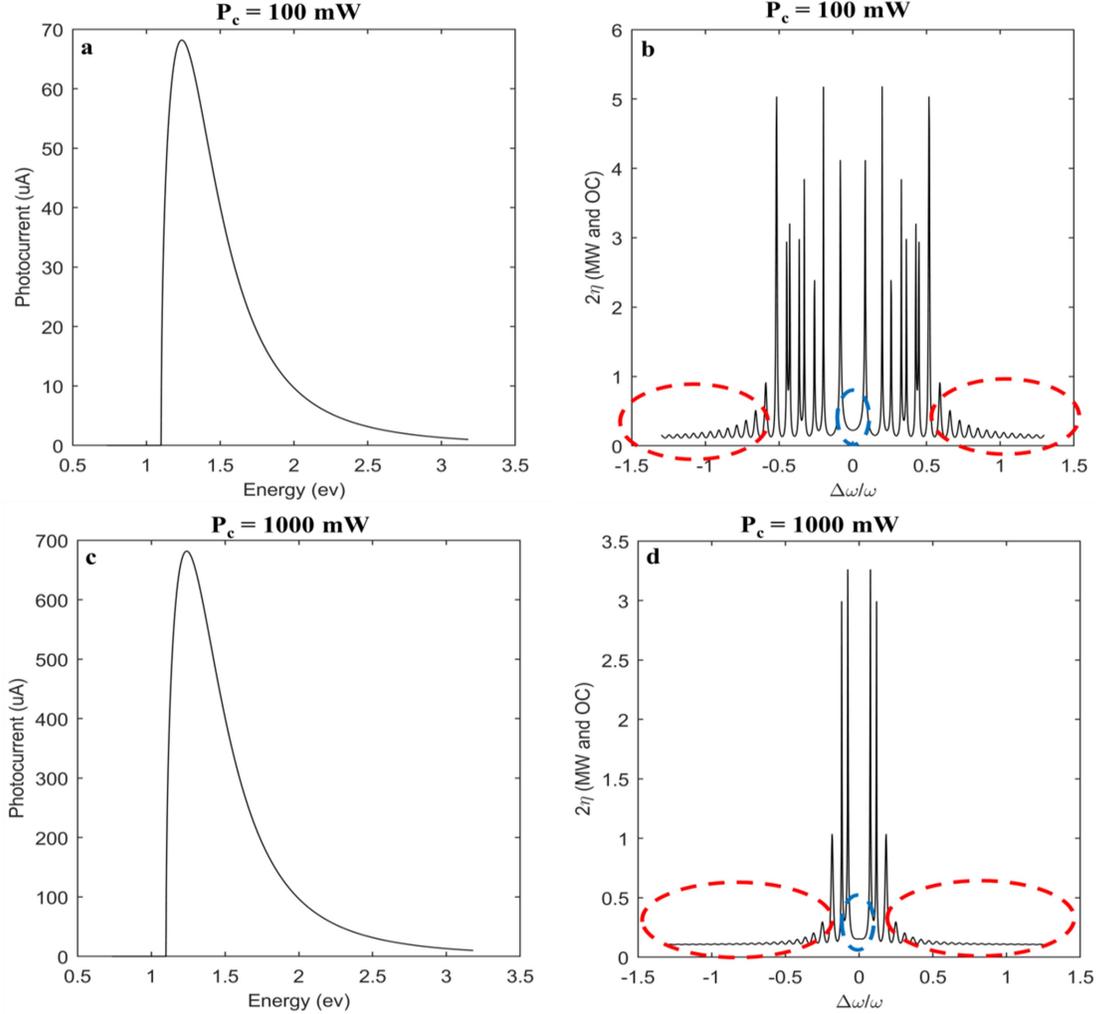

Fig. 5 The effect of the optical cavity pumping on the photocurrent (left handed) and entanglement between two modes (right handed) at different optical cavity pumping; (a, b) $P_c$ = 100 mW, (c, d) $P_c$ = 1000 mW; T = 298 K, $\Delta_c = 0.0\omega_\omega$; , $P_w$ = 10 mW.

Furthermore, in Fig. 3f, when the temperature is increased to 310 K, at $\Delta_\omega/\omega = 0$ entanglements is starting to be declined; however, $2\eta$ remains smaller than unity, and it is not a pure separable. We can briefly deduce that using the optoelectronic system rather than the traditional optomechanical system helps to reduce the system thermal induced noise, so the entangled modes are easily accessible at room temperature. Indeed, by selecting an optoelectronic element as a coupler between the OC and MW cavity, the thermal photon generating, which strongly disturbs the system operation is fixed. Hence, the mechanical element in the optomechanical system is a critical case for introducing thermal photon noise into a system that dramatically restricts the entanglement between modes at high temperatures. With knowledge of the optoelectronic system advantages, which were discussed above, one can briefly deduce that this type of system is the best case for replacement of the optomechanical system in quantum illumination systems. Moreover, by performing some supplementary modeling we showed that using optoelectronic system leads to manage at which detuning frequency the entanglement between cavity modes is maximized. The results are shown in Fig. 4. This figure reveals the entanglement between modes for different detuning factor $\Delta_c$ in which entanglement between modes tends to be maximum if and only if $\Delta_c = \Delta_\omega$. Therefore, by engineering the detuning factor, the entanglement center frequency, which indicates that two modes are fully entangled, is changed. To our knowledge, this point can be considered as a degree of freedom in quantum properties engineering or more specific in the entanglement engineering. From these simulations, one can briefly deduce that by controlling the OC modes coupling to the photodiode, the flowing current is changed, and thus the Varactor diode biased voltage is mutually altered. This effect simultaneously manipulates the MW cavity modes. In fact, by photocurrent controlling, the coupled modes entangling will be engineered. One of the methods for easily manipulating the entangling between modes is the OC pumping amplitude ($P_c$). From Fig. 1d, it is deduced that by changing the $P_c$ amplitude, the photocurrent amplitude is altered. In fact, this is the case that directly manipulates the Varactor diode's biased voltage, and then the MW cavity mode is alterable. For this reason, a few simulations were done in which the effect of the OC pumping amplitude changing is considered. Initially, we considered the effect of $P_c$ on the photocurrent amplitude. It is clear that by increasing the OC driving power, the generated current amplitude $A_c$ is increased, which also schematically illustrated in Fig. 1d. For better understanding, one shall compare the Fig. 5a and c. This can be clearly proved by Eq. 13. However, the more interesting case occurs in the entanglement between two modes. With a quick glance, it is comprehensible that by $P_c$ increasing, the amplitude of the 2η generally decreases, which means that by increasing the coupling factor, the modes are going to be strongly entangled. In the following, comparing Fig. 3e with Fig. 5b and 5d reveals that the amount of the divergences in the entanglement profile for different $\Delta_\omega$ values is dramatically reduced. This decline can be explained by the fact that by increasing the $P_c$, the total transition rate (Fig. 5c) is increased; so, it is proven that by intensifying the coupling factor, the entanglement between modes is broadened for large $\Delta_\omega$ (red dashed line). Furthermore, as one of the interesting points of this work, entanglement at $\Delta_\omega = 0$ is not affected (blue dashed line).

### IV. Conclusions

In this article, a novel optoelectronics system is designed to couple the optical cavity mode to the microwave cavity mode. The original aim was to investigate the entanglement between the OC and MW cavity modes. For this reason, the dynamics of the motion of the designed system was analyzed with Heisenberg-Langevin equations. The main objective of this work is to generate the photocurrent by the photodetector, which is directly coupled to the OC; which was theoretically proved by the perturbation theory. Next, it is examined that the photocurrent can affect the Varactor diode bias voltage, suggesting that the Varactor internal capacitance is directly dependent on the photocurrent. In fact, we studied the feasibility of coupling the optical modes to the microcavity modes through the photocurrent and Varactor diode capacitor. Finally, we analyzed the entanglement between OC and MW cavity modes and found that two modes were entangled at room temperature. Besides, via the modeling results, it was shown that the traditional optomechanical system could be substituted by the novel optoelectronic system. Moreover, the optoelectronic system rather than the optomechanical one gives a degree of freedom to manage the output modes entanglement, which allows reducing the complexities due to the mechanical elements.